\documentclass[pre,aps,11pt]{revtex4-1} 

\usepackage{xcolor}
\usepackage{amsmath}  
\usepackage{amsfonts} 
\usepackage{graphicx} 
\usepackage[utf8]{inputenc}
\usepackage{gensymb}
\begin{document}


\title{Free surface of a liquid in a rotating frame with time-depend velocity
}

\author{Martín Monteiro}
\email{monteiro@ort.edu.uy} 
\affiliation{Universidad ORT Uruguay}

\author{Fernando Tornaría}
\email{ftornaria@gmail.com}

\affiliation{CES-ANEP, Montevideo,  Uruguay}

\author{Arturo C. Martí}
\email{marti@fisica.edu.uy}

\affiliation{Instituto de F\'{i}sica, Facultad de Ciencias, Universidad de la Rep\'{u}blica,
             Igu\'{a} 4225, Montevideo, 11200, Uruguay}


\date{\today}

\begin{abstract}

The shape of liquid surface in a rotating frame depends on the angular
velocity. In this experiment, a fluid in a rectangular container with
a small width is placed on
a rotating table. A smartphone fixed to the
rotating frame simultaneously records the fluid surface with the
camera and also, thanks to the built-in gyroscope, the angular
velocity.  When the table starts rotating the surface evolves and develops a parabolic shape. Using
video analysis we obtain the surface's shape:
concavity of the parabole
and height of the vertex. Experimental results are
compared with theoretical predictions. This problem contributes to
improve the understanding of relevant concepts in fluid dynamics.

\end{abstract}

\maketitle 

\section{Statement of the problem}
When we gently stir a cup of coffee  or tea the free surface develops a familiar
parabolic shape. This in fact an expression of an ubiquitous phenomenon called vortex.
In general, in fluid mechanics, vortex motion, is characterized by fluid elements moving
along circular streamlines \cite{whitefluid}. 
There are two basic types of vortex flows:
irrotational vortex, that occurs tipically around a sink, and rotational vortex or 
solid body rotation, that occurs in the aforementioned example.
In particular, the free surface of a rotating liquid
develops the well-known parabolic profile whose characteristics depends on
the angular velocity. This last is an usual problem in introductory
courses when dealing with fluid mechanics.
Although it is not difficult from the
theoretical point of view, it is not easy to address it
experimentally. Here we propose an experiment to analyze the parabolic
shape of a liquid surface in a rotating table as a function of the
time-dependent angular velocity.

Rotating fluids were studied in several experiments. To mention a few, the profile of
circular uniform motion of liquid surface was determined using a
vertical laser beam reflected from the curved surface
\cite{graf1997apparatus} and determine the acceleration of
gravity. More recently \cite{sundstrom2016measuring}, this
experimental setup was improved using the fact that a rotating liquid
surface will form a parabolic reflector which will focus light into a
unique focal point. Another interesting experiment is the Newton’s
bucket which provides a simple demonstration that simulates Mach’s
principle allowing to observe the concave shape of the liquid
\cite{de2016rotating}. In other experiments rotating fluids were
studied in the framework of the equivalence principle and non-inertial
frames \cite{fagerlind2015liquid,tornaria2014understanding}.

In the present experiment a narrow container is placed on a rotating
table whose angular velocity can be manually controlled. As shown in
the next Section the liquid surface develops a parabolic shape whose
concavity and the location of the vertex can be related to the angular
velocity of the table
and to the gravitational acceleration. The experimental setup, described in Section
\ref{sec:experiment}, in additon to the container on the rotating
table, includes a smartphone, also fixed to the rotating table, allows
us to register the shape of the liquid surface with the camera and the
angular velocity with the gyroscope. This ability to measure
simultaneously with more than one sensor is a great advantage of
smartphones since it allows us to perform a great variety of
experiments, even outdoors, avoiding the dependence on fragile or
unavailable instruments (see for example Refs.
\cite{Monteiro2014angular,Monteiro2014exploring,monteiro2016exploring,monteiro2017polarization,monteiro2017magnetic}). Thanks
to the analysis of the digital video it is possible to readily obtain
the characteristic of the parabolic shape. The results are presented
in Section \ref{sec:res} and, finally, the conclusion is given in
Section \ref{sec:con}.

\section{Shape of a liquid surface in a rotating frame}
\label{sec:theo}

The free surface of a liquid in a rotating frame is obtained from the
points where the pressure is equal to the atmospheric pressure.  Let
us consider the pressure field in a fluid, $p(\vec{r})$, subjected to
a constant acceleration $\vec{a}$ and a gravitational field
$\vec{g}$. After transients, when a fluid is rotating as a rigid body,
\textit{i.e.} the fluid elements follow circular streamlines without
deforming and viscous stresses are null \cite{whitefluid}.  Under
these hypothesis, pressure gradient, gravitational field and particle
acceleration are related by the simple expression
\begin{equation}
 \nabla p = \rho (\vec{g} -\vec{a}).
\label{eq:pressue-stationary}
\end{equation}

\begin{figure}[h]
\includegraphics[width=.9\columnwidth]{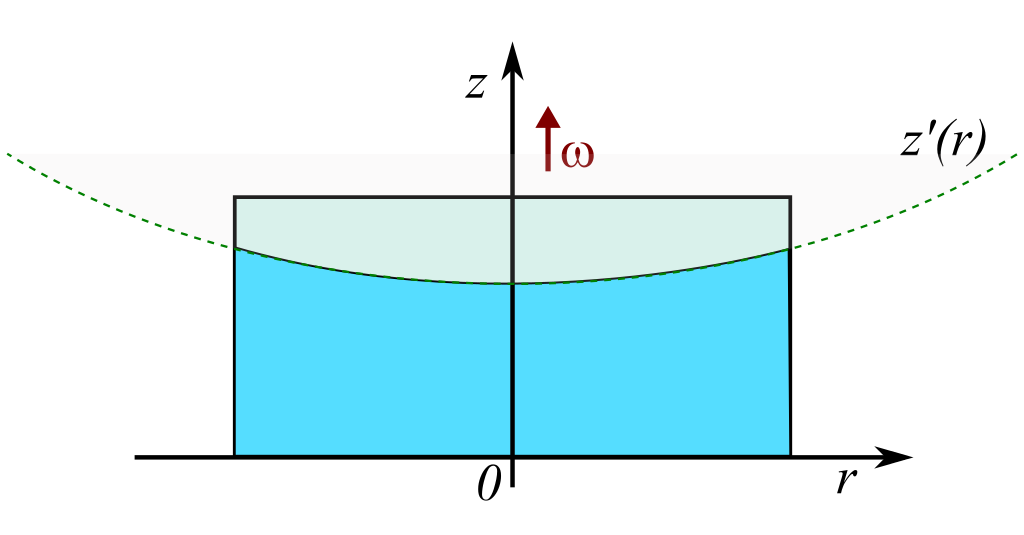}
\caption{A liquid in a prismatic container with a free surface, $z'(r)$, mounted
  on a rotating table with angular velocity $\omega$ displays a
  parabolic shape. The figure also indicates the definition of the
  coordinate axes in the relative system.}
\label{figtanque}
\end{figure}

In the present experiment, we consider a fluid in a narrow prismatic
container, as shown in Fig.~\ref{figtanque}, whose basis is $L \times
d$ where $L \gg d$ and its height is large enough so that the fluid
does not overflow.  When the system is at rest, the fluid, with
density $\rho$ and negligible viscosity, reaches a height $H$. The container is placed on a rotating table whose angular velocity, $\omega$, around the vertical axis 
passing through the geometrical center
can be externally controlled. In this experiment
the angular velocity is slowly varied so that the
transient effects can be neglectd.
Figure~\ref{figtanque} also displays the cylindrical polar
coordinates with unitary vectors $(\hat{r},\hat{\theta},\hat{z})$,
where $\hat{r}$ coincides with the basis of the container and
$\hat{z}$ is a vertical axis through the center of the container.

Under these assumptions, the velocity field is that 
of a rigid body and can be expressed as
$\vec{u}= \omega r \hat{\theta}$ while the acceleration $\vec{a}= -
\omega^2 r \hat{r}$.  To obtain the pressure field, after substituting
these expressions in the Eq.~(\ref{eq:pressue-stationary}) we obtain
\begin{equation}
- \omega^2 r  \hat{r} =  - \frac{\nabla p}{\rho} - g \hat{z}.
\label{eq:euler-stationary}
\end{equation}
where in the case of an axisymmetric field the gradient can be written as
\begin{equation}
 \nabla p=\frac {\partial p}{ \partial r}  \hat{r} 
+ \frac {\partial p}{ \partial z}  \hat{z}.
\end{equation}
The pressure field  can be easily integrated 
to obtain
\begin{equation}
p(r,z) = -  \rho g z + \frac{\rho  \omega^2 r^2}{2} + C
\label{eq:pressure}
\end{equation}
where $C$ is a constant of integration with dimensions of pressure.
The equation of the free  surface, $z'(r)$, is obtained using the
constraint that the pressure corresponds to the atmospheric pressure
$p_{atm}$ and results
\begin{equation}
z'(r) =\frac{r^2 \omega^2}{2 g} - \frac{p_{atm}}{\rho g} + \frac{C}{\rho g}.
\label{eq:surface}
\end{equation}
The constant $C$ can be obtained using the mass conservation and the
fact that the fluid is incompressible,
\begin{equation}
HL = 2 \int_0^{L/2}{z'(r)dr} = 2 \int_0^{L/2}{\left( \frac{r^2
    \omega^2}{2 g} - \frac{p_{atm}}{\rho g} + \frac{C}{\rho g}\right)
}dr.
 \label{eq:conservation}
\end{equation}
Performing the integral we get the expression for $C$
\begin{equation}
C= p_{atm}  + \rho g H - \frac{\rho \omega^2L^2}{12}.
 \label{eq:C}
\end{equation}
Finally, the pressure field inside the fluid can be expressed as
\begin{equation}
p(r,z) = p_{atm} + \rho g(H- z) + \frac{\rho \omega^2} {2} \left(r^2-
\frac{L^2}{24} \right)
\label{eq:prz}
\end{equation}
where we can appreciate static and dynamics contributions.
The free surface can be finally expressed as
\begin{equation}
z'(r) =H  - \frac{\omega^2}{2 g} \left( \frac{L^2}{12}-r^2\right)
\label{eq:surf2}
\end{equation}

We notice that the concavity and the location of the vertex of the
parabole depend on the angular velocity. The vertex of the parabole,
given by $r=0$ is located at 
\begin{equation}
z_v =H - \frac{\omega^2 L^2}{24 g}. 
\label{eqzv}
\end{equation}
When
the angular velocity is $ \omega \ge \sqrt{24 g H}/L$ the parabole
vertex reaches the bottom of the container and these expressions are
not longer valid. It is also interesting that there are two
\textit{nodal} points given by $z'(r_0)=H$ with $r_0= \pm L
/\sqrt{12}$ that always belong to the free surface.

\section{Experimental set-up and data processing}
\label{sec:experiment}

\begin{figure}[h]
\centering
\includegraphics[width=.8\columnwidth]{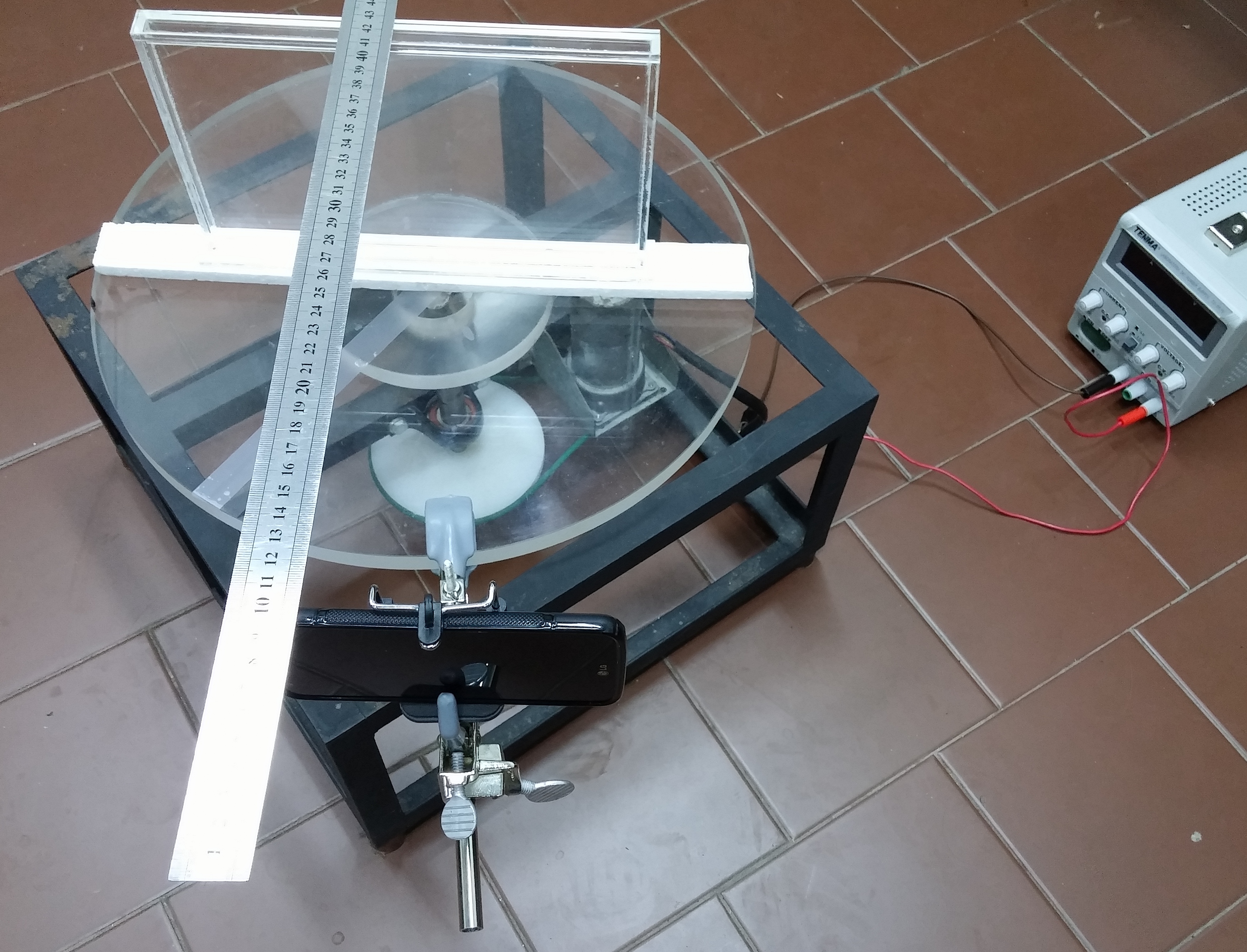}
\caption{Experimental set-up composed of a narrow container mounted on
  a rotatory table. The smartphone, also fixed to the rotating system,
  supplies both the video of the of the time-evolving surface and the
  angular velocity obtained with the gyroscope.}
\label{figsetup}
\end{figure}

The experimental setup shown in Fig.~\ref{figsetup} consists of a prismatic container and a smartphone, model LG-G2
(with digital camera and built-in gyroscope), both of them fixed to a rotating table. The dimensions of the container containing dyed water were $25$ cm width, $15$ cm height and $2$ cm thickness. The rotating table  was powered by a DC motor so the rotational speed could be adjusted by varying the voltage applied to the motor.

Initially, with the rotating table  at rest we turn on the video camera and  we start recording the  measures provided with the gyroscope and proximity
sensors with the Androsensor \textit{app}. To synchronize the video and the and data provided by the \textit{app}, we cover a few seconds simultaneously the lens of the camera and the proximity sensor. In this way we obtain a common time reference for the video and sensor data.

\begin{figure}[h!]
\centering
\includegraphics[width=.9\columnwidth]{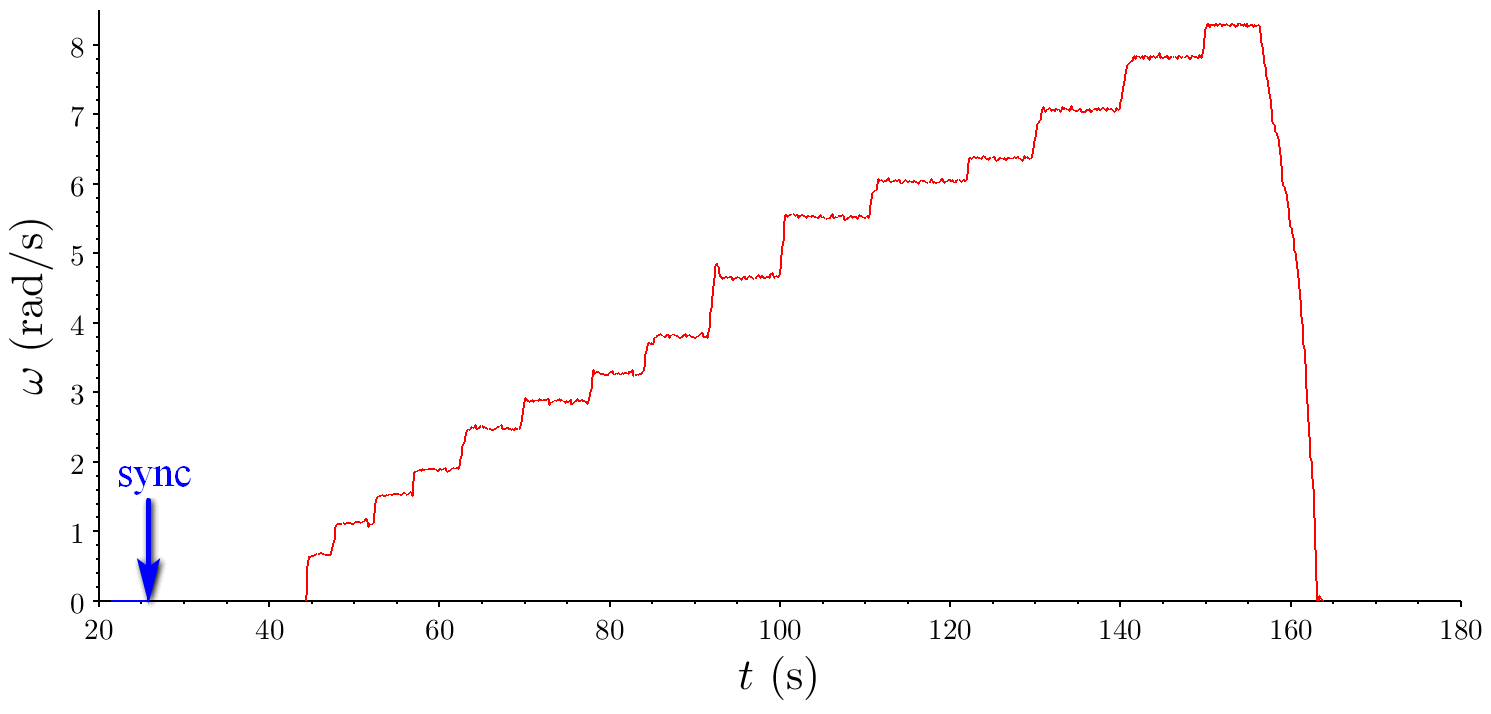}
\caption{Temporal evolution of the angular velocity. The appreciable
  jumps are produced by the operator regulating the DC power
  supply. 
  The blue arrow indicates the instant in which the hand uncover the camera and
  the proximity sensor to register a mark to synchronize the video and the gyroscope sensor.
  }
\label{fig:omega}
\end{figure}

Then, we  turn on the power supply and the rotating table starts rotating.
Throughout the experiment the power supply is slowly increased by small 
jumps and so does the angular velocity. Figure \ref{fig:omega} shows the temporal evolution of the angular velocity. The blue arrow indicates the time used to synchronize the video and the sensor data. By the end of the experiment both files, video and sensor data, are transferred to a computer. \textcolor{red}{The accompanying video abstract shows a synopsis of the experimental video.
The video is available at
 https://youtu.be/6SsX16rNoVE.
}

To analyze the characteristics of the time-evolving surface, firstly, 
we extract the individual frames from the digital video. Next, we select
15 frames, corresponding to different values of the angular velocities 
displayed in Fig.~\ref{fig:omega}. Each frame was analyzed with the video analysis software Tracker \cite{brown2009tracker}.  Several points, tipically 8,
on the interface were manually labeled and, then, we perform
a parabolic fit, $y = Ax^2+Bx+C$. The coefficient $A$ corresponds to the
concavity of the parabole. From the coefficients $B$ and $C$ the height
of the vertex, $H = -B^2/(4A)$, can be readily obtained.

\begin{figure}[h!]
\centering
\includegraphics[width=.9\columnwidth]{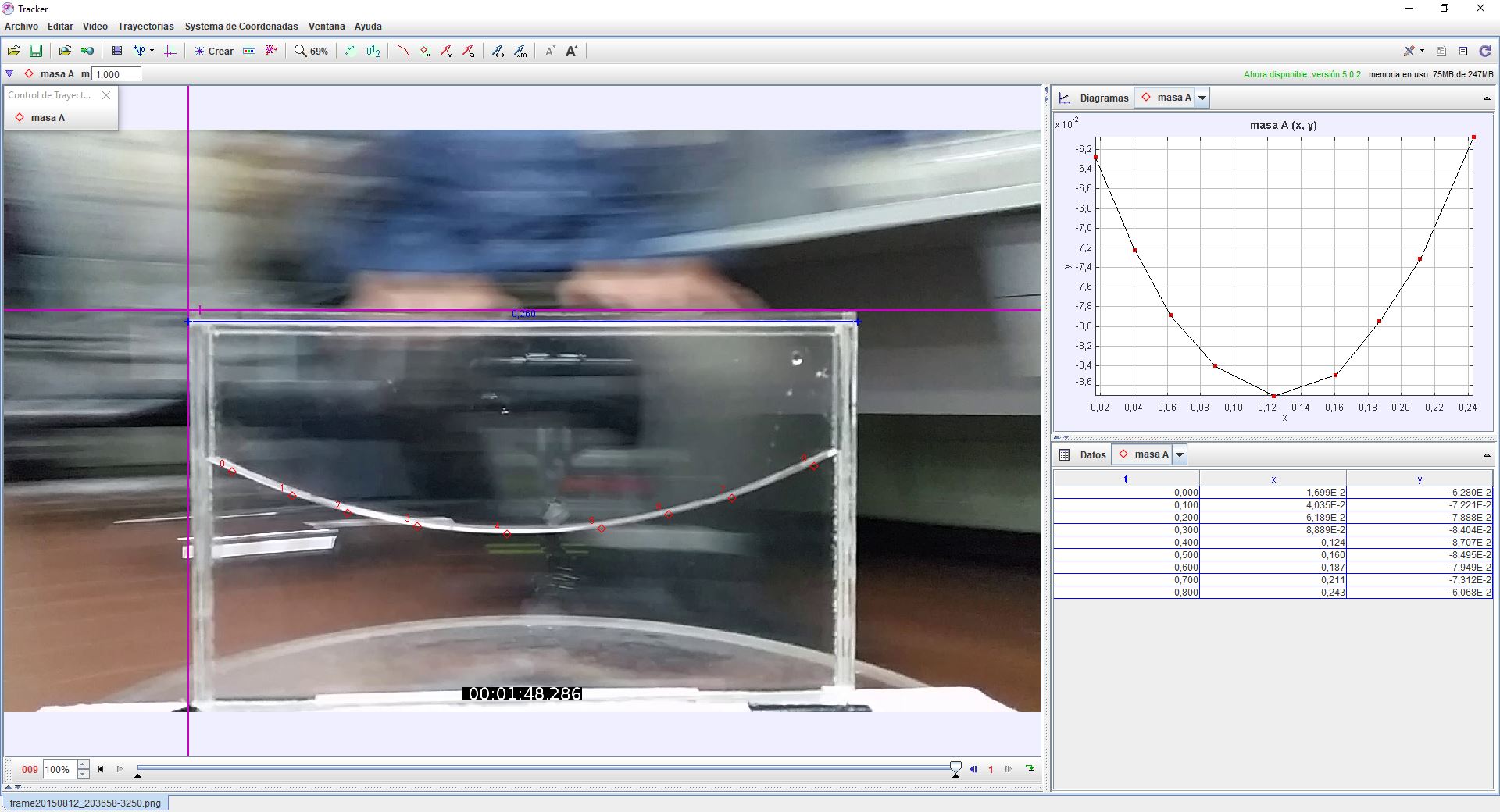}
\caption{Tracker screen-shot showing one frame of the digital video with the free surface and the  points selected (left). The right panel shows the parabolic shape
and the coordinates of the selected points.}
\label{figtracker}
\end{figure}

\section{Results}
\label{sec:res}

\begin{figure}[h]
\centering
\includegraphics[width=.95\columnwidth]{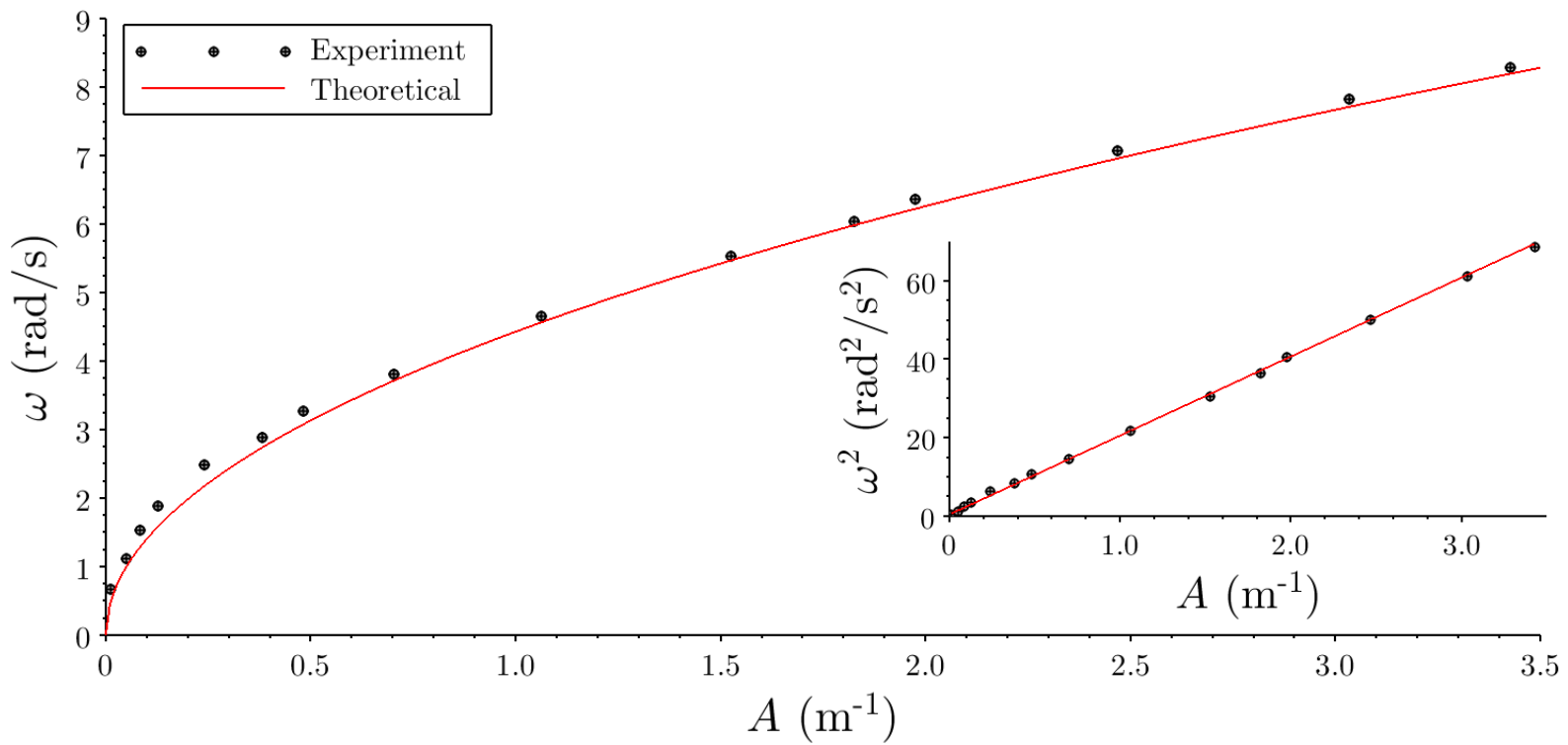}
\caption{Relationship between the angular velocity and the concavity
of the fluid surface fitted to a parabole. The points indicate the experimental
results and the solid line the model prediction. The slope of the linear fit 
shown in inset is $20.16 (4)$ m·rad$^2$/s$^2$.}
\label{fig02}
\end{figure}

After the data were processed, we compared the results for the concavity and
the height of the parabole with the model prediction. Figure~\ref{fig02} shows the experimental  relationship between the concavity of the parabole and the angular velocity and the model prediction. The slope of the linear fit  $20.16 (4)$ m·rad$^2$/s$^2$, according to Eq.~\ref{eq:surf2}, is with good agreement $2g$. 

\begin{figure}[h]
\centering
\includegraphics[width=.95\columnwidth]{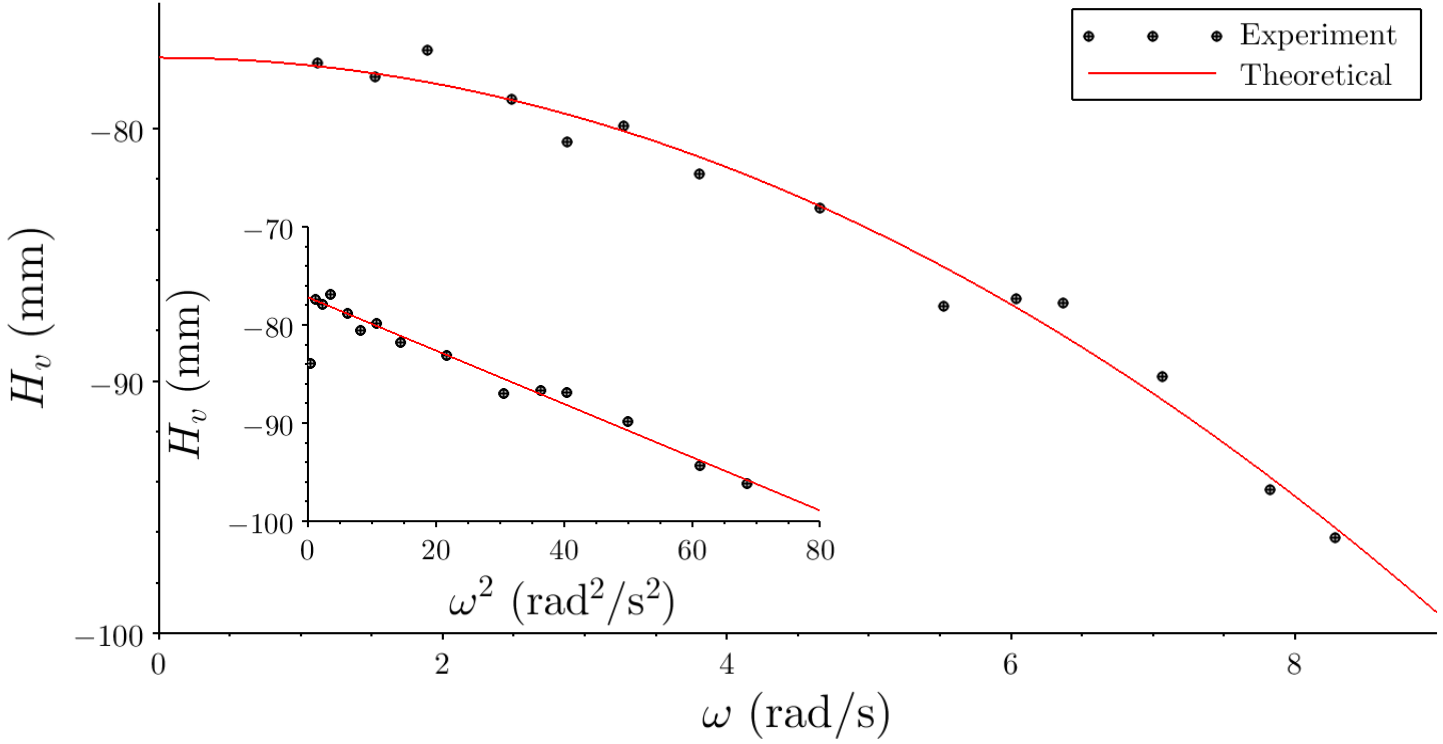}
\caption{Height of the parabole vertex as a function of the angular
  velocity squared. The red line indicates a linear fit (slope  $ -0.27(1)$ mm·s$^2$/rad$^2$ and intercept  $-7.72(3)$) . The leftmost
  point which corresponds to a small angular velocity (and therefore
  present a great uncertainty) was not taken into account in the
  linear fit shown in the inset.}
\label{figheight}
\end{figure}

The height of the parabole vertex as a function angular velocity squared is plotted in Figure~\ref{figheight}. The slope of the linear fit results $ -0.27(1)$ mm·s$^2$/rad$^2$ which is very similar to  the value, according
to Eq.~\ref{eqzv},  given by the model $ -L^2/24 g = -0.2655(5)$ mm·s$^2$/rad$^2$. In addition, the intercept corresponds to the water level with the rotating table at rest. In the experiment, the value obtained  is $-7.72(3)$cm while the direct value obtained measuring directly on the image is $ -7.6(2)$ cm.

\section{Conclusions and Prospects}
\label{sec:con}

The present proposal aims at experimenting with the free surface of a liquid rotating with a time dependent angular velocity. Thanks to a smartphone both the shape of the surface and the angular velocity are simultaneously measured. Using video analysis
software we obtain the coefficients of the parabolic profile which can be related to the angular velocity, the gravitational acceleration and the water level with the
rotating table at rest. 
The present experiment yields very good agreement with the theoretical model. This simple and inexpensive proposal provides an opportunity for students to engage with challenging
aspects of fluid dynamics without sophisticated or expensive equipment.

\begin{acknowledgments}
 We are very grateful to Cecilia Cabeza for productive discussions.
 This work was partially supported by the program \textit{Física nolineal, CSIC Grupos I+D} (Udelar, 
 Uruguay). 
\end{acknowledgments}

\bibliography{/home/arturo/Dropbox/bibtex/mybib}

\bibliographystyle{unsrt}

\end{document}